\documentclass[12pt]{article}
\usepackage{amssymb}
\usepackage{amsmath}
\usepackage{cite}

\numberwithin{equation}{section}

\newcommand{\be}{\begin{equation}}
\newcommand{\ee}{\end{equation}}
\newcommand{\non}{\nonumber}

\newcommand{\A}{\mathbb{A}}
\newcommand{\B}{\mathbb{B}}
\newcommand{\C}{\mathbb{C}}
\newcommand{\D}{\mathbb{D}}

\newcommand{\M}{\mathbb{M}}

\newcommand{\R}{\mathbb{R}}
\newcommand{\T}{\mathbb{T}}
\newcommand{\U}{\mathbb{U}}
\newcommand{\tr}{\mathop{\rm tr}\nolimits}
\newcommand{\diag}{\mathop{\rm diag}\nolimits}

\begin{document}

\begin{titlepage}
\strut\hfill UMTG--304
\vspace{.5in}
\begin{center}

\LARGE Q-systems with boundary parameters\\ 
\vspace{1in}
\large Rafael I. Nepomechie\footnote{nepomechie@miami.edu}\\
Physics Department, P.O. Box 248046\\
University of Miami, Coral Gables, FL 33124\\[0.8in]
\end{center}

\vspace{.5in}

\begin{abstract}
Q-systems provide an efficient way of solving Bethe equations.  We
formulate here Q-systems for both the isotropic and anisotropic open
Heisenberg quantum spin-1/2 chains with diagonal boundary magnetic
fields.  We check these Q-systems using novel discrete Wronskian-type formulas
(relating the fundamental Q-function and its dual) that involve the
boundary parameters.
\end{abstract}

\end{titlepage}

\setcounter{footnote}{0}

\section{Introduction}\label{sec:intro}

An efficient way of solving rational Bethe equations for periodic
models was introduced in 2016 by Marboe and Volin \cite{Marboe:2016yyn}.  This
so-called Q-system method, which is an outgrowth of a long line of research 
(see e.g. \cite{Krichever:1996qd, Kazakov:2007fy, Kuniba:2010ir} and 
references therein), has already been exploited in various
investigations, see e.g. \cite{Marboe:2017dmb, Basso:2017khq,
Suzuki:2017ipd, Ryan:2018fyo, Coronado:2018ypq, Jacobsen:2018pjt}.
This method was recently generalized \cite{Bajnok:2019zub} for the
rank-1 case in two different directions: from rational to
trigonometric, and from periodic to open boundary conditions.
However, there the models with boundaries were restricted to ``free''
boundary conditions, without any boundary parameters.  In this paper,
we show how to further generalize the rank-1 Q-system, so as to
incorporate two arbitrary boundary parameters.  To check the new
Q-systems, we use novel discrete Wronskian-type formulas (relating the
fundamental Q-function and its dual) that involve the boundary
parameters.

We start with the simpler case of the isotropic open Heisenberg
quantum spin-1/2 chain with boundary magnetic fields in Sec.
\ref{sec:rational}, and we then consider the anisotropic case in Sec.
\ref{sec:trig}.  We end with a brief conclusion in Sec.
\ref{sec:conclusion}.  If the boundary parameters are on a special manifold,
then a separate treatment of the Bethe equations is required, which is
presented in Appendix \ref{sec:special}.  The closed spin chain with
diagonal twisted boundary conditions is briefly discussed in Appendix
\ref{sec:transform}.

\section{Rational case}\label{sec:rational}

We consider here the isotropic (XXX) open Heisenberg quantum spin-1/2
chain of length $N$ with boundary magnetic fields, whose Hamiltonian
is given by
\be
H = \sum_{k=1}^{N-1} \vec \sigma_{k} \cdot  \vec \sigma_{k+1}
- \frac{1}{\beta} \sigma^{z}_{1} + \frac{1}{\alpha} \sigma^{z}_{N} \,, 
\label{Hamrat}
\ee
where $\alpha$ and $\beta$ are arbitrary 
parameters.\footnote{If we require $H$ to be Hermitian, then $\alpha$ 
and $\beta$ must be real.} This model 
is $U(1)$ invariant
\be
\left[ H \,, S^{z} \right] = 0 \,, \qquad S^{z} = \sum_{k=1}^{N} 
\frac{1}{2}\sigma^{z} \,.
\label{U1}
\ee
The special rational case considered in \cite{Bajnok:2019zub} corresponds to the limit where 
both $\alpha$ and $\beta$ tend to infinity, in which case the model 
becomes $SU(2)$ invariant.
We discuss the Bethe ansatz solution of this model in Sec. \ref{sec:BArat}, and we 
present corresponding Q-systems in Sec. \ref{sec:QQrat}.

\subsection{Bethe ansatz}\label{sec:BArat}

We begin by briefly reviewing the algebraic Bethe ansatz solution of
the model (\ref{Hamrat}) in Sec. \ref{sec:ABArat}. The dual Bethe 
equations and dual TQ-equations are presented in Sec. \ref{sec:dualrat}. 
Starting from a Q-function, we show in Sec. \ref{sec:Prat} 
how to construct the corresponding dual Q-function, which will be needed for 
our later discussion of the Q-system. We shall see that the Q-function and its dual
are related by a novel Wronskian-type formula that involves the boundary
parameters.

\subsubsection{Algebraic Bethe ansatz}\label{sec:ABArat}

The algebraic Bethe ansatz solution of
the model (\ref{Hamrat}) was formulated by Sklyanin 
\cite{Sklyanin:1988yz}.\footnote{The coordinate Bethe ansatz solution 
was found in \cite{Gaudin:1971zza, Alcaraz:1987uk}. For an 
introduction to algebraic Bethe ansatz, see e.g. \cite{Faddeev:1996iy}.}  
Following the notations in \cite{Bajnok:2019zub}, we consider the R-matrix (solution of the
Yang-Baxter equation) given by the $4 \times 4$ matrix
\be
\mathbb{R}(u)=(u-\frac{i}{2})\mathbb{I}+i\mathbb{P}\,,
\label{Rmatrat}
\ee
where $\mathbb{P}$ is the permutation matrix and $\mathbb{I}$ is the 
identity matrix. We define the monodromy matrices 
\begin{align}
\mathbb{M}_{0}(u) &=\mathbb{R}_{01}(u)\, 
\mathbb{R}_{02}(u)\dots\mathbb{R}_{0N}(u) \,, \non \\
\widehat{\M}_{0}(u) &= \R_{0 N}(u) \cdots \R_{0 2}(u)\, \R_{0 1}(u) \,.
\end{align}
We consider the K-matrices (solutions of boundary Yang-Baxter 
equations) given by the diagonal $2 \times 2$ matrices
\begin{align}
\mathbb{K}^{R}(u) &=\diag\left(i(\alpha-\tfrac{1}{2}) + u\,,  
i(\alpha+\tfrac{1}{2}) - u\right) \,, \non \\
\mathbb{K}^{L}(u) &=\diag\left(i(\beta-\tfrac{1}{2}) - u\,,  
i(\beta+\tfrac{1}{2}) + u \right) \,,
\label{Krat}
\end{align}
which evidently depend on the boundary parameters $\alpha$ and 
$\beta$, respectively.
The transfer matrix $\T(u) = \T(u; \alpha, \beta)$ is given by \cite{Sklyanin:1988yz}
\be
\T(u) = \tr_{0}  \U_{0}(u) \,, \qquad 
\U_{0}(u) = \mathbb{K}^{L}_{0}(u)\, \M_{0}(u)\, \mathbb{K}^{R}_{0}(u)\, \widehat{\M}_{0}(u) 
\,,
\label{transfrat}
\ee
which has the commutativity property
\be
\left[ \T(u) \,, \T(v) \right] = 0 
\ee
and satisfies $ \T(-u) =  \T(u)$.
The Hamiltonian (\ref{Hamrat}) is equal to 
$\frac{(-1)^{N}}{2i\alpha\beta}\frac{d\T(u)}{du}\Big\vert_{u=i/2}$, 
up to an additive constant. If either of the boundary parameters 
vanishes, then a local integrable Hamiltonian can be obtained from the 
second derivative of the transfer matrix at $u=i/2$.

In order to construct eigenstates of the transfer matrix,
we define the operators $\A(u)$, $\B(u)$, $\C(u)$, $\D(u)$ from 
$\U_{0}(u)$ (\ref{transfrat}) as follows\footnote{This construction is similar 
to, but not the same as, the one in \cite{Sklyanin:1988yz}.}
\be
\U_{0}(u) = \left( \begin{array}{cc}
\delta_{0}(u)\, \A(u) & \B(u) \\
\C(u) & \delta_{1}(u)\, \D(u) + \delta_{2}(u)\, \A(u) 
\end{array} \right) \,,
\ee 
with 
\be
\delta_{0}(u) = \frac{u- i(\beta-\frac{1}{2})}{u- 
i(\beta+\frac{1}{2})}\,, \qquad 
\delta_{1}(u) = 
\frac{(u-\frac{i}{2})}{u} \,, 
\qquad
\delta_{2}(u) =  -\frac{i 
(u+i(\beta+\frac{1}{2}))}{2u(u-i(\beta+\frac{1}{2}))} \,.
\qquad
\ee
Choosing the reference state
\be
|0\rangle = {1\choose 0}^{\otimes N} \,,
\label{reference}
\ee
the Bethe states are defined by
\be
|u_{1} \ldots u_{M} \rangle = \prod_{k=1}^{M} \B(u_{k})  |0\rangle \,.
\label{Bethestaterat} 
\ee
These states are eigenstates of the transfer matrix $\T(u)$ with 
eigenvalues $T(u)$
\be
\T(u) |u_{1} \ldots u_{M} \rangle = T(u) 
|u_{1} \ldots u_{M} \rangle \,,
\ee
provided that $\{ u_{1}\,, \ldots \,, u_{M} \}$ are admissible 
solutions of the Bethe equations \footnote{Some solutions of the 
Bethe equations do not correspond to eigenvalues and eigenvectors of the 
transfer matrix. Examples include solutions with repeated Bethe 
roots; and, since the spin chain is open, solutions with the Bethe roots 0 or $\pm i/2$, see 
\cite{Bajnok:2019zub} and references therein. Here we 
call \emph{admissible} those solutions of the Bethe equations that \emph{do}
correspond to genuine eigenvalues and eigenvectors of the transfer matrix.}
\begin{align}
\frac{g(u_{j}-\tfrac{i}{2})}{f(u_{j}+\tfrac{i}{2})}
\left(\frac{u_{j}+\tfrac{i}{2}}{u_{j}-\tfrac{i}{2}}\right)^{2N}
& =\prod_{k=1; k \ne j}^{M} 
\frac{(u_{j}-u_{k}+i)(u_{j}+u_{k}+i)}{(u_{j}-u_{k}-i)(u_{j}+u_{k}-i)}
\,, \non \\
&\qquad\qquad j = 1, \ldots, M \,, \qquad M = 0, \ldots, N \,,
\label{BAErat}
\end{align}
where we have introduced the functions $f(u)$ and $g(u)$ defined by
\be
f(u) = (u- i \alpha)(u + i \beta) \,, \qquad g(u) = f(-u) = (u+ i 
\alpha)(u - i \beta) \,,
\label{fgrat}
\ee
which will play an important role in the following. 
The eigenvalues $T(u)$ (which are necessarily polynomials in $u^{2}$) are given by the TQ-equation
\be
- u\, T(u)\, Q(u) = (u^{+})^{2N+1} g^{-}(u)\, Q^{--}(u) + (u^{-})^{2N+1} 
f^{+}(u)\, Q^{++}(u) \,,
\label{TQrat}
\ee
where $Q(u)$ is also a polynomial in $u^{2}$ defined by
\be
Q(u) = \prod_{k=1}^{M} (u-u_{k})(u+u_{k}) \,,
\label{Qrat}
\ee
and we use the standard notation $F^{\pm}(u) = F(u \pm 
\tfrac{i}{2})$ for any function $F(u)$. For generic values of the 
boundary parameters, the transfer-matrix eigenvalues $T(u)$ are not 
degenerate.  For the special case $\alpha-\beta=1$, the functions
in (\ref{fgrat}) satisfy $f^{+} = g^{-}$, and the Bethe equations
(\ref{BAErat}) must be slightly modified.  This special case is discussed
further in Appendix \ref{sec:special}.

\subsubsection{Duality}\label{sec:dualrat}

We observe that the transfer matrix is not invariant under charge 
conjugation ${\cal C}= (\sigma^{x})^{\otimes N}$; indeed, 
the boundary parameters become negated
\be
{\cal C}\, \T(u; \alpha, \beta) \, {\cal C} = \T(u; -\alpha, -\beta) 
\,.
\ee
Similarly, the $\B$ and $\C$ operators are related by
\be
{\cal C}\, \B(u; \alpha, \beta) \, {\cal C} = \C(u; -\alpha, -\beta) 
\label{BCrat}
\,.
\ee

A given eigenstate of the transfer matrix with eigenvalue $T(u)$
can be represented either by a Bethe state (\ref{Bethestaterat}), 
or by a corresponding ``dual'' Bethe state
\be
|\tilde{u}_{1} \ldots \tilde{u}_{\tilde{M}} \rangle = 
\prod_{k=1}^{\tilde{M}} \C(\tilde{u}_{k})  |\tilde{0}\rangle 
\label{dualBethestaterat} 
\ee
constructed with the  ``dual'' reference state
\be
|\tilde{0}\rangle = {\cal C}\,|0\rangle = {0\choose 1}^{\otimes N} \,.
\label{dualreference}
\ee
Hence, the dual Bethe state satisfies
\be
\T(u) |\tilde{u}_{1} \ldots \tilde{u}_{\tilde{M}} \rangle = T(u) 
|\tilde{u}_{1} \ldots \tilde{u}_{\tilde{M}} \rangle \,,
\ee
where $\{ \tilde{u}_{1}\,, \ldots \,, \tilde{u}_{\tilde{M}} \}$ satisfy the 
``dual'' Bethe equations\footnote{This fact was already noticed in 
\cite{Sklyanin:1988yz}.}
\begin{align}
\frac{f(\tilde{u}_{j}-\tfrac{i}{2})}{g(\tilde{u}_{j}+\tfrac{i}{2})}
\left(\frac{\tilde{u}_{j}+\tfrac{i}{2}}{\tilde{u}_{j}-\tfrac{i}{2}}\right)^{2N}
& =\prod_{k=1; k \ne j}^{\tilde{M}} 
\frac{(\tilde{u}_{j}-\tilde{u}_{k}+i)(\tilde{u}_{j}+\tilde{u}_{k}+i)}
{(\tilde{u}_{j}-\tilde{u}_{k}-i)(\tilde{u}_{j}+\tilde{u}_{k}-i)}
\,, \non \\
&\qquad\qquad j = 1, \ldots, \tilde{M} \,, \qquad \tilde{M} = 0, \ldots, N \,.
\label{BAEratdual}
\end{align}
In terms of the dual Bethe roots, the 
eigenvalues $T(u)$ are given by a ``dual'' TQ-equation
\be
- u\, T(u)\, P(u) = (u^{+})^{2N+1} f^{-}(u)\, P^{--}(u) + (u^{-})^{2N+1} 
g^{+}(u)\, P^{++}(u) \,,
\label{TQratdual}
\ee
where $P(u)$ is the corresponding ``dual'' Q-function\footnote{We do not 
specify the overall constant in 
(\ref{Prat1}), which will be specified in (\ref{Wronskrat}) below.}
\be
P(u) \propto \prod_{k=1}^{\tilde{M}} 
(u-\tilde{u}_{k})(u+\tilde{u}_{k}) \,.
\label{Prat1}
\ee
(The symbol $\propto$ used here and below denotes proportionality, 
i.e. equality up to a constant.)
The Bethe equations (\ref{BAErat}) and their duals 
(\ref{BAEratdual}), and similarly the TQ equations (\ref{TQrat}) 
and (\ref{TQratdual}), are related (as follows from (\ref{BCrat}))
by $\alpha \mapsto -\alpha$ and $\beta \mapsto -\beta$, together 
with $u_{j} \mapsto \tilde{u}_{j}$. The fact that the Bethe equations and TQ-equation are 
\emph{not} self-dual is a new feature of this problem. 
Indeed, for the periodic closed chain and for the $SU(2)$-invariant open 
chain considered earlier \cite{Bajnok:2019zub}, the Bethe equations 
and TQ-equation \emph{are} self-dual in this sense.

\subsubsection{Wronskian-type formula}\label{sec:Prat}

We now show that $Q(u)$ and $P(u)$ (the dual Q-function) satisfy a
Wronskian-type formula, which will be needed in Sec.  \ref{sec:QQrat}.
The result is similar to the one for the periodic closed chain
\cite{Pronko:1998xa}; however, there are some interesting new
features, since the TQ-equation is no longer self-dual.

We begin by multiplying both sides of the TQ-equation (\ref{TQrat}) by $P$, 
\be
- u\, T\, Q\, P = (u^{+})^{2N+1} g^{-}\, Q^{--}\, P + (u^{-})^{2N+1} 
f^{+}\, Q^{++}\, P \,;
\label{TQratP}
\ee
and multiplying both sides of the dual TQ-equation (\ref{TQratdual}) 
by $Q$, 
\be
- u\, T\, Q\, P = (u^{+})^{2N+1} f^{-}\, P^{--}\, Q + (u^{-})^{2N+1} 
g^{+}\, P^{++}\, Q \,.
\label{TQratdualQ}
\ee
Subtracting these two equations, we obtain
\be
0 = (u^{+})^{2N+1} W^{-} - (u^{-})^{2N+1} W^{+} \,,
\label{Wfuncreqn}
\ee
where we have introduced the function $W(u)$ defined by
\be
W = g\, P^{+}\, Q^{-} - f\, P^{-}\, Q^{+} \,.
\label{Wdef}
\ee
We regard (\ref{Wfuncreqn}) as a functional equation
for $W$, which has the solution $W 
\propto u^{2N+1}$. Since the proportionality constant is generically 
nonzero (exceptions can occur on the special 
manifold $\alpha-\beta=\pm 1$, see (\ref{zeroW})), we normalize $P$ so that the 
constant is 1. We conclude that 
\be
g\, P^{+}\, Q^{-} - f\, P^{-}\, Q^{+} = u\, Q_{0,0} \,,
\label{Wronskrat}
\ee
where we have set
\be
Q_{0,0}(u) = u^{2N}
\label{Q00ratant}
\ee
in anticipation of a notation that will be introduced in Sec. \ref{sec:QQrat}. 
The result (\ref{Wronskrat}) is an important discrete Wronskian-type formula relating $Q$ and $P$, 
which interestingly is ``deformed'' by the functions $f$ and $g$ (\ref{fgrat}).
Note that 
\be
M + \tilde{M} = N \,,
\ee
where $2M$ and $2\tilde{M}$ are the degrees of $Q$ and $P$, 
respectively. We also note that by using (\ref{Wronskrat}) to eliminate $(u^{\pm})^{2N+1}$ 
from the TQ-equation (\ref{TQrat}), we obtain 
\be
- u\, T = g^{+}\, g^{-}\, P^{++}\, Q^{--} - f^{+}\, f^{-}\, P^{--}\, 
Q^{++} \,,
\label{TPQrat}
\ee
which is a deformation of another well-known result \cite{Pronko:1998xa}.

Let us pause to underscore the new insight that this problem has
revealed.  In the context of quantum integrability, the discrete Wronskian (or
Casoratian) formula has generally been regarded (see e.g.
\cite{Pronko:1998xa}) as a relation between two solutions of the
\emph{same} finite-difference TQ-equation.  However, we now recognize
this to be an exceptional situation, which occurs when the TQ-equation
is self-dual.  We should instead regard the Wronskian formula as a
relation between a solution of the TQ-equation and a solution of the
dual TQ-equation; and these two TQ-equations are generally \emph{not}
the same.\footnote{In fact, the TQ-equation (\ref{TQrat}) has only one
polynomial solution, instead of two.  This can be understood
heuristically from the fact that, in contrast with
\cite{Pronko:1998xa}, here there is no notion of ``equator'': due to
the absence of $SU(2)$ symmetry, it is necessary to include values of
$M$ (the number of Bethe roots) up to $N$, instead of $N/2$. Indeed, 
a second solution of (\ref{TQrat}) is $P' = U P$, where $U$ satisfies
$\frac{U^{+}}{U^{-}} = \frac{g}{f}$, which has a solution in terms of 
products of gamma functions
\be
U(u) = \Gamma(i u + \tfrac{1}{2} +\alpha)\, \Gamma(-i u + 
\tfrac{1}{2} + \alpha)\, \Gamma(i u + \tfrac{1}{2} - 
\beta)\, \Gamma(-i u + \tfrac{1}{2} - \beta) \,, \non
\ee
see Appendix \ref{sec:transform}.}

For the periodic chain, it was proved in \cite{Mukhin:2009,
Tarasov:2018} that polynomiality of $P(u)$ is equivalent to the
admissibility of the solution $\{u_{1}, \ldots, u_{M}\}$ of the Bethe
equations.  We conjecture that a similar result is true for the open spin
chain with diagonal boundary fields; namely, only those solutions $\{ u_{1}, 
\ldots, u_{M} \}$ of the Bethe equations (\ref{BAErat})
for which there exist corresponding solutions
$\{ \tilde{u}_{1}, \ldots, \tilde{u}_{\tilde{M}} \}$ of the dual 
Bethe equations (\ref{BAEratdual})
such that the Wronskian formula (\ref{Wronskrat}) is satisfied are admissible.

\subsection{Q-systems}\label{sec:QQrat}

We now look for a Q-system for the Bethe equations 
(\ref{BAErat}). Surprisingly, the answer is not unique; and we 
present two such Q-systems. For both systems,
we take (as anticipated in (\ref{Q00ratant}))
\be
Q_{0,0}(u) = u^{2N} \,,
\label{Q00rat}
\ee
and we identify $Q_{1,0}(u)$ as the fundamental Q-function (\ref{Qrat})
\be
Q_{1,0}(u) = Q(u) = \prod_{k=1}^{M} (u-u_{k})(u+u_{k}) 
= \sum_{k=0}^{M-1} c_{k}\, u^{2k} + u^{2M} \,.
\ee
For the rank-1 case that we consider in this paper, there are only two 
nontrivial sets of Q-functions, namely, $Q_{0,n}(u)$ and $Q_{1,n}(u)$. Our 
two Q-systems are distinguished by which of these two sets of 
Q-functions are ``deformed'' by the functions $f$ and $g$ defined in (\ref{fgrat}). 
We consider these systems separately in 
Secs. \ref{sec:Irat} and \ref{sec:IIrat}.

\subsubsection{Deforming $Q_{1,n}$}\label{sec:Irat}

We begin by considering the following Q-system
\begin{align}
u\, Q_{1,n} & \propto f^{[-(n-1)]}\, Q_{1,n-1}^{+} - g^{[n-1]}\, 
Q_{1,n-1}^{-} \,, \qquad n=1, 2,\ldots \,, \non \\
u\, Q_{0,n}\, Q_{1,n-1} & \propto Q_{1,n}^{+}\, Q_{0,n-1}^{-} - 
Q_{1,n}^{-}\, Q_{0,n-1}^{+} \,, \qquad\qquad n=1, 2, \ldots \,, \label{QQIrat}
\end{align}
where we use the notation $F^{[n]}(u) = F(u + \frac{n i}{2})$, and we 
remind the reader that the functions $f$ and $g$ are defined in (\ref{fgrat}). 
Comparing with the corresponding Q-system in \cite{Bajnok:2019zub}, we see 
that only the relations for $Q_{1,n}$ are deformed by $f$ and $g$. We 
claim that all of these Q-functions are polynomials if and only if $\{ 
u_{1}, \ldots, u_{M}\}$ (given by zeros of $Q_{1,0}(u)$) is an admissible solution of the Bethe 
equations (\ref{BAErat}).

As a preliminary check of this Q-system, let us verify that it leads to 
the Bethe equations (\ref{BAErat}). Eqs. (\ref{QQIrat}) with $n=1$ 
read
\begin{align}
u\, Q_{1,1}(u)  & \propto f(u)\, Q_{1,0}^{+}(u) - g(u)\, 
Q_{1,0}^{-}(u) \,, \label{QQI1rat} \\
u\, Q_{0,1}(u)\, Q_{1,0}(u) & \propto Q_{1,1}^{+}(u)\, Q_{0,0}^{-}(u) - 
Q_{1,1}^{-}(u)\, Q_{0,0}^{+}(u) \,. \label{QQI0rat} 
\end{align}
Performing in (\ref{QQI1rat}) the shifts $u \mapsto u \pm \frac{i}{2}$ and evaluating at a Bethe root 
$u=u_{j}$, we obtain
\be
(u_{j} +\tfrac{i}{2})\,  Q_{1,1}^{+}(u_{j}) \propto 
f^{+}(u_{j})\, Q^{++}(u_{j}) \,, \qquad 
(u_{j} -\tfrac{i}{2})\,  Q_{1,1}^{-}(u_{j}) \propto 
- g^{-}(u_{j})\, Q^{--}(u_{j})  \,,
\label{QQstep1}
\ee
since $Q(u_{j}) = 0$. Evaluating (\ref{QQI0rat}) at $u=u_{j}$ gives
\be
Q_{1,1}^{+}(u_{j})\, Q_{0,0}^{-}(u_{j}) = Q_{1,1}^{-}(u_{j})\, Q_{0,0}^{+}(u_{j}) \,.
\label{QQstep2}
\ee
Finally, substituting (\ref{QQstep1}) into (\ref{QQstep2}), we indeed arrive 
at the Bethe equations (\ref{BAErat}). We have also  
verified numerically for small values of $N$ that this Q-system reproduces the 
complete spectrum of the transfer matrix.

We can solve the Q-system (\ref{QQIrat}) in terms of $Q(u)$ and a 
function $P(u)$ defined by the Wronskian-type formula (\ref{Wronskrat}). 
Indeed, we find that this system is solved by \footnote{Somewhat similar formulas have been known in 
the context of supersymmetric spin chains \cite{Tsuboi:2009ud, 
Tsuboi:2011iz}.}
\begin{align}
Q_{1,n} & \propto D^{n}Q 	\,, \non \\
u\, Q_{0,n} & \propto g^{[n]}\, (D^{n}P)^{+}\, (D^{n}Q)^{-} - 
f^{[-n]}\, (D^{n}P)^{-}\, (D^{n}Q)^{+} \,,
\label{solIrat}
\end{align}
where, as in \cite{Bajnok:2019zub}, $D^{n}P$ is defined by
\begin{align}
D^{n}P &= \frac{1}{u}\left[(D^{n-1}P)^{+} - (D^{n-1}P)^{-} \right] 
\,, \qquad n= 1, 2, \ldots \,,
\label{DPIrat}
\end{align}
(with $D^{0}=1$) and we now define $D^{n}Q$ by
\begin{align}
D^{n}Q &= \frac{1}{u}\left[f^{[-(n-1)]}\, (D^{n-1}Q)^{+} - 
g^{[n-1]}\,(D^{n-1}Q)^{-} \right]  \qquad n= 1, 2, \ldots  \,.
\label{DQIrat}
\end{align}
Note that the expression for $D^{n}P$ (\ref{DPIrat}) is not deformed, while the 
expression for $D^{n}Q$ (\ref{DQIrat}) is deformed by $f$ and $g$.
The solution (\ref{solIrat}) shows that polynomiality of $P(u)$ is 
equivalent to polynomiality of all the Q-functions. Since
polynomiality of $P(u)$ is equivalent to the admissibility of the 
solution $\{u_{1}, \ldots, u_{M}\}$ (see Sec. \ref{sec:Prat}), we 
conclude that (\ref{QQIrat}) is indeed a Q-system for the model 
(\ref{Hamrat}).

\subsubsection{Deforming $Q_{0,n}$}\label{sec:IIrat}

We now consider a different Q-system
\begin{align}
u\, Q_{1,n} & \propto Q_{1,n-1}^{+} - Q_{1,n-1}^{-} \,, \qquad\qquad\qquad\qquad\qquad n=1, 2,\ldots \,, \non \\
u\, Q_{0,n}\, Q_{1,n-1} & \propto f^{[n]}\, Q_{1,n}^{+}\, Q_{0,n-1}^{-} - 
g^{[-n]}\, Q_{1,n}^{-}\, Q_{0,n-1}^{+} \,, \qquad n=1, 2, \ldots \,, 
\label{QQIIrat}
\end{align}
where now the relations for $Q_{0,n}$ are deformed by $f$ and $g$.  We 
again claim that all the Q-functions are polynomials if and only if $\{ 
u_{1}, \ldots, u_{M}\}$ is an admissible solution of the Bethe 
equations (\ref{BAErat}).

Let us begin by verifying that this Q-system also leads to 
the Bethe equations (\ref{BAErat}). Eqs. (\ref{QQIIrat}) with $n=1$ 
read
\begin{align}
u\, Q_{1,1}(u)  & \propto  Q_{1,0}^{+}(u) - Q_{1,0}^{-}(u) \,, 
\label{QQII1rat} \\
u\, Q_{0,1}(u)\, Q_{1,0}(u) & \propto f^{+}(u)\, Q_{1,1}^{+}(u)\, Q_{0,0}^{-}(u) - 
g^{-}(u)\, Q_{1,1}^{-}(u)\, Q_{0,0}^{+}(u) \,. \label{QQII0rat} 
\end{align}
Performing in (\ref{QQII1rat}) the shifts $u \mapsto u \pm \frac{i}{2}$ and evaluating at a Bethe root 
$u=u_{j}$, we obtain
\be
(u_{j} +\tfrac{i}{2})\,  Q_{1,1}^{+}(u_{j}) \propto 
 Q^{++}(u_{j}) \,, \qquad 
(u_{j} -\tfrac{i}{2})\,  Q_{1,1}^{-}(u_{j}) \propto 
- Q^{--}(u_{j})  \,.
\label{QQIIstep1}
\ee
Evaluating (\ref{QQII0rat}) at $u=u_{j}$ gives
\be
f^{+}(u_{j})\, Q_{1,1}^{+}(u_{j})\, Q_{0,0}^{-}(u_{j}) = g^{-}(u_{j})\, Q_{1,1}^{-}(u_{j})\, Q_{0,0}^{+}(u_{j}) \,.
\label{QQIIstep2}
\ee
Finally, substituting (\ref{QQIIstep1}) into (\ref{QQIIstep2}), we  
again arrive at the Bethe equations (\ref{BAErat}). We have also 
verified numerically for small values of $N$ that this Q-system reproduces the 
complete spectrum of the transfer matrix.

We can also solve the Q-system (\ref{QQIIrat}) in terms of $Q(u)$ and a 
function $P(u)$ defined by the Wronskian-type formula (\ref{Wronskrat}). 
Indeed, we find that this system is solved by
\begin{align}
Q_{1,n} & \propto D^{n}Q 	\,, \non \\
u\, Q_{0,n} & \propto g^{[-n]}\, (D^{n}P)^{+}\, (D^{n}Q)^{-} - 
f^{[n]}\, (D^{n}P)^{-}\, (D^{n}Q)^{+} \,,
\label{solIIrat}
\end{align}
where $D^{n}P$ is now defined by
\begin{align}
D^{n}P &= \frac{1}{u}\left[g^{[-(n-1)]}\, (D^{n-1}P)^{+} - f^{[n-1]}\, (D^{n-1}P)^{-} \right] 
\,, \qquad n= 1, 2, \ldots \,,
\label{DPIIrat}
\end{align}
while $D^{n}Q$ is defined by
\begin{align}
D^{n}Q &= \frac{1}{u}\left[(D^{n-1}Q)^{+} - 
(D^{n-1}Q)^{-} \right]  \qquad n= 1, 2, \ldots  \,.
\label{DQIIrat}
\end{align}
Note that now the expression for $D^{n}P$ (\ref{DPIIrat}) is deformed, while the 
expression for $D^{n}Q$ (\ref{DQIIrat}) is not deformed.
The solution (\ref{solIIrat}) shows that polynomiality of $P(u)$ is 
equivalent to polynomiality of all the Q-functions, hence 
(\ref{QQIIrat}) is also a Q-system for the model (\ref{Hamrat}).

\section{Trigonometric case}\label{sec:trig}

We now consider the anisotropic (XXZ) open Heisenberg quantum 
spin-1/2 chain of length $N$ with anisotropy parameter $\eta$ and with diagonal
boundary magnetic fields, whose Hamiltonian is given by
\be
H = \sum_{k=1}^{N-1} \left[ \sigma^{x}_{k} \sigma^{x}_{k+1} + 
\sigma^{y}_{k} \sigma^{y}_{k+1} + \cosh(\eta) \sigma^{z}_{k} 
\sigma^{z}_{k+1} \right] 
- \sinh(\eta)\coth(\beta \eta) \sigma^{z}_{1} + 
\sinh(\eta)\coth(\alpha \eta)\sigma^{z}_{N} \,,
\label{Hamtrig}
\ee
where $\alpha$ and $\beta$ are arbitrary parameters. This model reduces to the isotropic model (\ref{Hamrat})
in the limit $\eta \rightarrow 0$; and, like the latter, is $U(1)$ invariant (\ref{U1}).
The special trigonometric case considered in \cite{Bajnok:2019zub} corresponds to the limit where 
both $\alpha$ and $\beta$ tend to infinity, in which case the model 
becomes $U_{q}(su(2))$ invariant.
We first discuss the Bethe ansatz solution of this model in Sec. \ref{sec:BAtrig}, and we 
then present corresponding Q-systems in Sec. \ref{sec:QQtrig}.

\subsection{Bethe ansatz}\label{sec:BAtrig}

The algebraic Bethe ansatz solution of the anisotropic model (\ref{Hamtrig}) is 
similar to the one for its isotropic limit (\ref{Hamrat}) discussed 
in Sec. \ref{sec:BArat}. Hence, we present only the salient formulas.

The R-matrix is now given by
\be
\mathbb{R}(u)=\left(\begin{array}{cccc}
\sinh(u+\frac{\eta}{2}) & 0 & 0 & 0\\
0 & \sinh(u-\frac{\eta}{2}) & \sinh(\eta) & 0\\
0 & \sinh(\eta) & \sinh(u-\frac{\eta}{2}) & 0\\
0 & 0 & 0 & \sinh(u+\frac{\eta}{2}) 
\end{array}\right) \,,
\label{Rmattrig}
\end{equation}
and the K-matrices are given by
\begin{align}
\mathbb{K}^{L}(u) &=\diag\left( \sinh(\eta(\alpha-\tfrac{1}{2}) + u)\,,  
\sinh(\eta(\alpha+\tfrac{1}{2}) - u)\right) \,, \non \\
\mathbb{K}^{R}(u) &=\diag\left( \sinh(\eta(\beta-\tfrac{1}{2}) - u)\,,  
\sinh(\eta(\beta+\tfrac{1}{2}) + u) \right) \,.
\label{Ktrig}
\end{align}
These matrices reduce to (\ref{Rmatrat}) and (\ref{Krat}), 
respectively, by setting $u \mapsto \epsilon u\,, \eta  \mapsto i 
\epsilon$ and letting $\epsilon \rightarrow 0$. The Hamiltonian 
(\ref{Hamtrig}) is proportional to 
$\frac{d\T(u)}{du}\Big\vert_{u=\eta/2}$, up to an additive constant. 

The Bethe equations are now given by
\begin{align}
\frac{g(u_{j}-\tfrac{\eta}{2})}{f(u_{j}+\tfrac{\eta}{2})}
\left(\frac{\sinh(u_{j}+\tfrac{\eta}{2})}{\sinh(u_{j}-\tfrac{\eta}{2})}\right)^{2N}
& =\prod_{k=1; k \ne j}^{M} 
\frac{\sinh(u_{j}-u_{k}+\eta)\, 
\sinh(u_{j}+u_{k}+\eta)}{\sinh(u_{j}-u_{k}-\eta)\, \sinh(u_{j}+u_{k}-\eta)}
\,, \non \\
&\qquad\qquad j = 1, \ldots, M \,, \qquad M = 0, \ldots, N \,,
\label{BAEtrig}
\end{align}
where the functions $f(u)$ and $g(u)$ are now given by
\be
f(u) = \sinh(u - \eta \alpha)\, \sinh(u + \eta \beta) \,, \qquad g(u) 
= f(-u) = \sinh(u + \eta \alpha)\, \sinh(u - \eta \beta) \,,
\label{fgtrig}
\ee
cf. (\ref{BAErat}), (\ref{fgrat}). The TQ-equation becomes
\begin{align}
- \sinh(2u)\, T(u)\, Q(u) &= \sinh(2u+\eta)\, 
\sinh^{2N}(u+\tfrac{\eta}{2})\, g^{-}(u)\, Q^{--}(u) \non \\
& + \sinh(2u-\eta)\, \sinh^{2N}(u-\tfrac{\eta}{2})\, f^{+}(u)\, Q^{++}(u) \,,
\label{TQtrig}
\end{align}
with
\be
Q(u) = \prod_{k=1}^{M} \sinh(u-u_{k})\, \sinh(u+u_{k}) \,,
\label{Qtrig}
\ee
where we now use the notation $F^{\pm}(u) = F(u \pm \tfrac{\eta}{2})$, cf. (\ref{TQrat}), 
(\ref{Qrat}). The Q-function is a polynomial in $t^{2}$ and $t^{-2}$, 
where $t=e^{u}$.

The dual Bethe equations are now given by
\begin{align}
\frac{f(\tilde{u}_{j}-\tfrac{\eta}{2})}{g(\tilde{u}_{j}+\tfrac{\eta}{2})}
\left(\frac{\sinh(\tilde{u}_{j}+\tfrac{\eta}{2})}{\sinh(\tilde{u}_{j}-\tfrac{\eta}{2})}\right)^{2N}
& =\prod_{k=1; k \ne j}^{\tilde{M}} 
\frac{\sinh(\tilde{u}_{j}-\tilde{u}_{k}+\eta)\, 
\sinh(\tilde{u}_{j}+\tilde{u}_{k}+\eta)}
{\sinh(\tilde{u}_{j}-\tilde{u}_{k}-\eta)\, \sinh(\tilde{u}_{j}+\tilde{u}_{k}-\eta)}
\,, \non \\
&\qquad\qquad j = 1, \ldots, \tilde{M} \,, \qquad \tilde{M} = 0, \ldots, N \,,
\label{BAEtrigdual}
\end{align}
and the dual TQ-equation is 
\begin{align}
- \sinh(2u)\, T(u)\, P(u) &= \sinh(2u+\eta)\, 
\sinh^{2N}(u+\tfrac{\eta}{2})\, f^{-}(u)\, P^{--}(u) \non \\
& + \sinh(2u-\eta)\, \sinh^{2N}(u-\tfrac{\eta}{2})\, g^{+}(u)\, P^{++}(u) \,,
\label{TQtrigdual}
\end{align}
where $P(u)$ is the corresponding dual Q-function
\be
P(u) \propto \prod_{k=1}^{\tilde{M}} \sinh(u-\tilde{u}_{k})\, \sinh(u+\tilde{u}_{k}) \,,
\label{Ptrig1}
\ee
cf. (\ref{BAEratdual})-(\ref{Prat1}). Finally, the Wronskian-type relation 
becomes
\be
g\, P^{+}\, Q^{-} - f\, P^{-}\, Q^{+} = \sinh(2u)\, Q_{0,0} \,,
\label{Wronsktrig}
\ee
where $Q_{0,0}(u)$ is given by (\ref{Q00trig}), cf. (\ref{TPQrat}). 
As in the rational case, more care is required when the boundary 
parameters lie on the special manifold $\alpha-\beta=\pm 1$.

\subsection{Q-systems}\label{sec:QQtrig}

We now look for a Q-system for the Bethe equations 
(\ref{BAEtrig}). As in the rational case, we find two such systems.
For both systems, we take 
\be
Q_{0,0}(u) = \sinh^{2N}(u) \,,
\label{Q00trig}
\ee
and we identify $Q_{1,0}(u)$ as the fundamental Q-function (\ref{Qtrig})
\be
Q_{1,0}(u) = Q(u) = \sum_{k=0}^{M-1} c_{k}\, (e^{2 u k} + e^{-2 u k}) 
+ e^{2 u M} + e^{-2 u M} \,.
\ee
We present the two Q-systems separately in 
Secs. \ref{sec:Itrig} and \ref{sec:IItrig}.

\subsubsection{Deforming $Q_{1,n}$}\label{sec:Itrig}

We begin by considering the following Q-system
\begin{align}
\sinh(2u)\, Q_{1,n} & \propto f^{[-(n-1)]}\, Q_{1,n-1}^{+} - g^{[n-1]}\, 
Q_{1,n-1}^{-} \,, \qquad n=1, 2,\ldots \,, \non \\
\sinh(2u)\, Q_{0,n}\, Q_{1,n-1} & \propto Q_{1,n}^{+}\, Q_{0,n-1}^{-} - 
Q_{1,n}^{-}\, Q_{0,n-1}^{+} \,, \qquad\qquad n=1, 2, \ldots \,, 
\label{QQItrig}
\end{align}
where now $F^{[n]}(u) = F(u + \frac{n \eta}{2})$, and 
the functions $f$ and $g$ are defined in (\ref{fgtrig}). 
As in (\ref{QQIrat}), the relations for $Q_{1,n}$ are deformed by $f$ and $g$.
All the Q-functions must now be polynomials in $t^{2}$ and $t^{-2}$, 
where $t=e^{u}$.

By repeating the steps (\ref{QQII0rat})-(\ref{QQIIstep2}), one can easily verify 
that the Q-system (\ref{QQItrig}) indeed leads to 
the Bethe equations (\ref{BAEtrig}). We have also 
verified numerically for small values of $N$ that this Q-system reproduces the 
complete spectrum of the transfer matrix.

We can solve the Q-system (\ref{QQItrig}) in terms of $Q(u)$ and a 
function $P(u)$ defined by the Wronskian-type formula (\ref{Wronsktrig}). 
Indeed, we find that this system is solved by
\begin{align}
Q_{1,n} & \propto D^{n}Q 	\,, \non \\
\sinh(2u)\, Q_{0,n} & \propto g^{[n]}\, (D^{n}P)^{+}\, (D^{n}Q)^{-} - 
f^{[-n]}\, (D^{n}P)^{-}\, (D^{n}Q)^{+} \,,
\label{solItrig}
\end{align}
where, similarly to \cite{Bajnok:2019zub}, $D^{n}P$ is defined by
\begin{align}
D^{n}P &= \frac{1}{\sinh(2u)}\left[(D^{n-1}P)^{+} - (D^{n-1}P)^{-} \right] 
\,, \qquad n= 1, 2, \ldots \,,
\label{DPItrig}
\end{align}
and we now define $D^{n}Q$ by
\begin{align}
D^{n}Q &= \frac{1}{\sinh(2u)}\left[f^{[-(n-1)]}\, (D^{n-1}Q)^{+} - 
g^{[n-1]}\,(D^{n-1}Q)^{-} \right]  \qquad n= 1, 2, \ldots  \,.
\label{DQItrig}
\end{align}
The solution (\ref{solItrig}) shows that polynomiality (in $t^{2}$ 
and $t^{-2}$) of $P(u)$ is equivalent to polynomiality of all the Q-functions, hence 
(\ref{QQItrig}) is indeed a Q-system for the model (\ref{Hamtrig}).

\subsubsection{Deforming $Q_{0,n}$}\label{sec:IItrig}

We now consider a different Q-system
\begin{align}
\sinh(2u)\, Q_{1,n} & \propto Q_{1,n-1}^{+} - Q_{1,n-1}^{-} \,, \qquad\qquad\qquad\qquad\qquad n=1, 2,\ldots \,, \non \\
\sinh(2u)\, Q_{0,n}\, Q_{1,n-1} & \propto f^{[n]}\, Q_{1,n}^{+}\, Q_{0,n-1}^{-} - 
g^{[-n]}\, Q_{1,n}^{-}\, Q_{0,n-1}^{+} \,, \qquad n=1, 2, \ldots \,, 
\label{QQIItrig}
\end{align}
where now the relations for $Q_{0,n}$ are deformed by $f$ and $g$. 
Again, all the Q-functions must be polynomials in $t^{2}$ and $t^{-2}$, 
where $t=e^{u}$.

By repeating the steps (\ref{QQII0rat})-(\ref{QQIIstep2}), we 
verify that this Q-system also leads to 
the Bethe equations (\ref{BAErat}). We have also 
verified numerically for small values of $N$ that this Q-system reproduces the 
complete spectrum of the transfer matrix.

We can also solve the Q-system (\ref{QQIItrig}) in terms of $Q(u)$ and a 
function $P(u)$ defined by the Wronskian-type relation (\ref{Wronsktrig}). 
Indeed, we find that this system is solved by
\begin{align}
Q_{1,n} & \propto D^{n}Q 	\,, \non \\
\sinh(2u)\, Q_{0,n} & \propto g^{[-n]}\, (D^{n}P)^{+}\, (D^{n}Q)^{-} - 
f^{[n]}\, (D^{n}P)^{-}\, (D^{n}Q)^{+} \,,
\label{solIItrig}
\end{align}
where $D^{n}P$ is now defined by
\begin{align}
D^{n}P &= \frac{1}{\sinh(2u)}\left[g^{[-(n-1)]}\, (D^{n-1}P)^{+} - f^{[n-1]}\, (D^{n-1}P)^{-} \right] 
\,, \qquad n= 1, 2, \ldots \,,
\label{DPIItrig}
\end{align}
while $D^{n}Q$ is defined by
\begin{align}
D^{n}Q &= \frac{1}{\sinh(2u)}\left[(D^{n-1}Q)^{+} - 
(D^{n-1}Q)^{-} \right]  \qquad n= 1, 2,  \ldots  \,.
\label{DQIItrig}
\end{align}
The solution (\ref{solIItrig}) shows that polynomiality (in $t^{2}$ 
and $t^{-2}$) of $P(u)$ is equivalent to polynomiality of all the Q-functions, hence 
(\ref{QQIItrig}) is also a Q-system for the model (\ref{Hamtrig}).

\section{Conclusions}\label{sec:conclusion}

We have shown that boundary parameters can be introduced in rank-1
Q-systems, for both the rational (\ref{QQIrat}), (\ref{QQIIrat}) and
trigonometric (\ref{QQItrig}), (\ref{QQIItrig}) cases.  We have also
found novel Wronskian-type formulas involving the boundary parameters
(\ref{Wronskrat}), (\ref{Wronsktrig}).  More generally, we have
recognized that such Wronskian formulas should be understood as
relations between a solution of the TQ-equation and a solution of the
dual TQ-equation; and that these two TQ-equations are generally \emph{not}
the same. We expect that these results will have
applications to various integrable boundary problems in AdS/CFT and
statistical mechanics, as has already occurred for integrable
periodic problems \cite{Marboe:2017dmb, Basso:2017khq, Suzuki:2017ipd,
Ryan:2018fyo, Coronado:2018ypq, Jacobsen:2018pjt}.

For the rational case, operators whose eigenvalues are given by $Q(u)$
and $P(u)$ have been constructed in \cite{Frassek:2015mra}, called 
there $Q_{+}$ and $Q_{-}$.  It would
be interesting to prove the Wronskian-type formula (\ref{QQIrat})
directly for the corresponding operators. These operators should  
provide, through (\ref{solIrat}) and  (\ref{solIIrat}), 
realizations of Q-operators satisfying the Q-systems (\ref{QQIrat}) 
and (\ref{QQIIrat}), respectively. Trigonometric
generalizations of these operators have been considered in
\cite{Baseilhac:2017hoz}.

We have restricted our attention here to cases with diagonal
K-matrices (\ref{Krat}), (\ref{Ktrig}).  It would be very interesting if these results could be
further generalized to cases with non-diagonal K-matrices 
\cite{Ghoshal:1993tm, deVega:1993xi}, where the
Bethe equations are significantly more complicated \cite{Cao:2013qxa,
Nepomechie:2013ila, Wang2015}.  It would also be interesting to
consider generalizations to rank higher than one. Indeed, perhaps we 
can now speculate that all integrable problems can be 
reformulated as Q-systems.

\section*{Acknowledgments}
I am grateful to Z. Bajnok, E. Granet and J. Jacobsen for their
collaboration on Q-systems for the special cases without boundary
parameters \cite{Bajnok:2019zub}. I also thank R. Frassek for 
bringing \cite{Frassek:2015mra, Baseilhac:2017hoz} to my attention,
and anonymous referees for their valuable comments.

\appendix

\section{The special manifold $\alpha-\beta=\pm 1$}\label{sec:special}

The Bethe equations (\ref{BAErat}) or dual Bethe equations
(\ref{BAEratdual}) become modified if the boundary parameters
$\alpha$ and $\beta$ are on the special manifold $\alpha-\beta=\pm 1$.
We consider separately the cases $\alpha-\beta=1$ and
$\alpha-\beta=-1$.

\subsection{The case $\alpha-\beta=1$}

We first consider the case when the boundary parameters satisfy 
$\alpha-\beta=1$. For definiteness, we set $\beta = \alpha -1$, with 
$\alpha$ a free parameter. For this case, since the 
functions in (\ref{fgrat}) satisfy $f^{+} = g^{-}$, the Bethe 
equations are no longer given by (\ref{BAErat}); instead, they are 
\begin{align}
f(u_{j}+\tfrac{i}{2})&\left[
\left(\frac{u_{j}+\tfrac{i}{2}}{u_{j}-\tfrac{i}{2}}\right)^{2N}
-\prod_{k=1; k \ne j}^{M} 
\frac{(u_{j}-u_{k}+i)(u_{j}+u_{k}+i)}{(u_{j}-u_{k}-i)(u_{j}+u_{k}-i)} 
\right]=0 \,, \non \\
&\qquad\qquad j = 1, \ldots, M \,, \qquad M = 0, \ldots, N \,.
\label{BAEratspecial}
\end{align}
Indeed, (\ref{BAEratspecial}) are the conditions for the RHS of the TQ-equation 
(\ref{TQrat}) to vanish when $f^{+} = g^{-}$ and $u=u_{j}$. In other 
words, a Bethe root $u_{j}$ must obey {\em either} 
$f(u_{j}+\tfrac{i}{2}) = 0$, i.e.
\be
u_{j} = \pm i (\alpha - \tfrac{1}{2}) \,,
\label{specialroot}
\ee
or 
\be
\left(\frac{u_{j}+\tfrac{i}{2}}{u_{j}-\tfrac{i}{2}}\right)^{2N}
=\prod_{k=1; k \ne j}^{M} 
\frac{(u_{j}-u_{k}+i)(u_{j}+u_{k}+i)}{(u_{j}-u_{k}-i)(u_{j}+u_{k}-i)} 
\label{BEsu2}
\,.
\ee
A similar example has been discussed in Appendix B.3 of 
\cite{Nepomechie:2017hgw}. Since $f^{-} \ne g^{+}$, the dual Bethe 
equations for this case are still given by (\ref{BAEratdual}).

Since Bethe roots are not repeated, there are only two possible classes of 
configurations of Bethe roots ${\cal S} = \{ u_{1}, \ldots, u_{M} \}$: 
\begin{description}
	\item[class I] ${\cal S}$ contains one Bethe root of the form 
	(\ref{specialroot})
	\item[class II] ${\cal S}$ does not contain any Bethe roots of the form 
	(\ref{specialroot})
\end{description}	

Let us first consider class I. Since the corresponding Q-function vanishes 
at $u=i (\alpha - \tfrac{1}{2})$, $Q(u)$ must contain $f^{+}(u)$ as 
one of its factors. We conjecture that
\be
Q(u) = f^{+}(u)\, P(u)  \qquad \mbox{(class I)} \,,
\label{QPconj}
\ee
where $P(u)$ is the dual Q-function. Indeed, by substituting 
(\ref{QPconj}) into the TQ-equation (\ref{TQrat}), and making use of 
the fact $f^{+} = g^{-}$, we obtain the dual TQ-equation 
(\ref{TQratdual}). Moreover, we have verified (\ref{QPconj}) 
numerically for several examples. It follows from (\ref{QPconj}) and 
$f^{+} = g^{-}$ that the Wronskian vanishes
\be
g\, P^{+} Q^{-} - f\, P^{-}\, Q^{+} = 0  \qquad \mbox{(class I)} \,.
\label{zeroW}
\ee

For class II, all the Bethe roots satisfy (\ref{BEsu2}), which are 
the Bethe equations for the $SU(2)$-invariant model (see e.g.
\cite{Bajnok:2019zub}). 

\subsection{The case $\alpha-\beta=-1$}

Let us now consider the case when the boundary parameters satisfy 
$\alpha-\beta=-1$. For definiteness, we set $\beta = \alpha +1$, with 
$\alpha$ a free parameter. Since $f^{+} \ne g^{-}$, the Bethe 
equations are still given by (\ref{BAErat}). However, now $f^{-} = 
g^{+}$; hence, the dual Bethe equations are no longer given by 
(\ref{BAEratdual}), and are instead given by
\begin{align}
f(\tilde{u}_{j}-\tfrac{i}{2})&\left[
\left(\frac{\tilde{u}_{j}+\tfrac{i}{2}}{\tilde{u}_{j}-\tfrac{i}{2}}\right)^{2N}
-\prod_{k=1; k \ne j}^{\tilde{M}} 
\frac{(\tilde{u}_{j}-\tilde{u}_{k}+i)(\tilde{u}_{j}+\tilde{u}_{k}+i)}
{(\tilde{u}_{j}-\tilde{u}_{k}-i)(\tilde{u}_{j}+\tilde{u}_{k}-i)} 
\right]=0\,, \non \\
&\qquad\qquad j = 1, \ldots, \tilde{M} \,, \qquad \tilde{M} = 0, \ldots, N \,.
\label{BAEratdualspecial}
\end{align}
These are the conditions for the RHS of the dual TQ-equation 
(\ref{TQratdual}) to vanish when $f^{-} = g^{+}$ and 
$u=\tilde{u}_{j}$.
Hence, a dual Bethe root $\tilde{u}_{j}$ must obey {\em either} 
$f(\tilde{u}_{j}-\tfrac{i}{2}) = 0$, i.e.
\be
\tilde{u}_{j} = \pm i (\alpha + \tfrac{1}{2}) \,,
\label{specialdualroot}
\ee
or 
\be
\left(\frac{\tilde{u}_{j}+\tfrac{i}{2}}{\tilde{u}_{j}-\tfrac{i}{2}}\right)^{2N}
=\prod_{k=1; k \ne j}^{\tilde{M}} 
\frac{(\tilde{u}_{j}-\tilde{u}_{k}+i)(\tilde{u}_{j}+\tilde{u}_{k}+i)}
{(\tilde{u}_{j}-\tilde{u}_{k}-i)(\tilde{u}_{j}+\tilde{u}_{k}-i)}
\label{BEsu2dual}
\,.
\ee 

There are two classes of configurations of dual Bethe roots 
$\tilde{\cal S} = \{ \tilde{u}_{1}, \ldots, \tilde{u}_{\tilde{M}} \}$: 
\begin{description}
	\item[class I] $\tilde{\cal S}$ contains one dual Bethe root of the form 
	(\ref{specialdualroot})
	\item[class II] $\tilde{\cal S}$ does not contain any dual Bethe roots of the form 
	(\ref{specialdualroot})
\end{description}	

For class I, since the dual Q-function vanishes 
at $u=i (\alpha + \tfrac{1}{2})$, $P(u)$ must contain $f^{-}(u)$ as 
one of its factors. We conjecture that
\be
P(u) = f^{-}(u)\, Q(u)  \qquad \mbox{(class I)} \,.
\label{QPconjdual}
\ee
Indeed, by substituting (\ref{QPconjdual}) into the dual TQ-equation
(\ref{TQratdual}), and making use of $f^{-} = g^{+}$, we obtain the
TQ-equation (\ref{TQrat}).  It follows that the Wronskian 
vanishes, as in (\ref{zeroW}).

\section{Transformations of TQ-equations and Wronskian 
relations}\label{sec:transform}

The open spin chain with diagonal boundary fields considered in this
paper is not the first example of a model whose TQ-equation and
its dual are not the same, and whose Wronskian relation does
not have the ordinary form.  Indeed, the closed chain with diagonal
twisted boundary conditions has long been known to also have these
features. Interestingly, for the latter model, it has also been known 
that there is a transformation that brings both the TQ-equation and 
its dual to the same ordinary form; and that there is a 
transformation that brings the Wronskian relation to the ordinary 
form. The price for obtaining the ordinary formulas is that the transformed 
functions (Q and/or P) are no longer polynomial. We briefly review 
these results in Sec. \ref{sec:twist}, and we then consider 
corresponding transformations for the open chain in Sec. \ref{sec:revisit}.

\subsection{Twisted boundary conditions}\label{sec:twist}

For the closed spin chain with diagonal twisted boundary conditions, the 
transfer matrix is given by
\be
\T(u) = \tr_{0}  \mathbb{F}_{0}\, \M_{0}(u) \,, \qquad 
\mathbb{F} = \diag(e^{-\frac{i\phi}{2}}\,, e^{\frac{i\phi}{2}})
\,,
\label{twisted}
\ee
where $\phi$ is a constant ($u$-independent) twist angle.  As is well
known, the corresponding TQ-equation is given by
\be
T Q  = e^{-\frac{i\phi}{2}} (u^{+})^{N} Q^{--} + e^{\frac{i\phi}{2}} (u^{-})^{N} 
Q^{++}  \,, \label{twistTQ}
\ee
the dual TQ-equation is given by
\be
T P  = e^{\frac{i\phi}{2}} (u^{+})^{N} P^{--} + e^{-\frac{i\phi}{2}} (u^{-})^{N} 
P^{++}  \,, \label{twistdualTQ}
\ee
while the discrete Wronskian relation is given by
\be
e^{-\frac{i\phi}{2}}  P^{+} Q^{-} - e^{\frac{i\phi}{2}}  P^{-} Q^{+}  = u^{N} \,.
\label{twistWronsk}
\ee 
Both $P(u)$ and $Q(u)$ are polynomials in $u$.
As in the case of the open chain with diagonal boundary fields discussed 
in Secs. \ref{sec:dualrat} and \ref{sec:Prat}, the 
TQ-equation (\ref{twistTQ}) is not the same as the dual TQ-equation 
(\ref{twistdualTQ}), and the 
Wronskian relation (\ref{twistWronsk}) is not the ordinary one. Nevertheless, 
as is also well known, it is 
possible to transform P and Q so that both transformed functions obey 
the same ordinary TQ-equation. Alternatively, by transforming 
only P, it is possible to bring the Wronskian relation to the ordinary form.

Indeed, in terms of the transformed functions
\be
Q'(u) = e^{\frac{u \phi}{2}}Q(u) \,, \qquad P'(u) = e^{-\frac{u 
\phi}{2}}P(u) \,,
\label{twisttransf1}
\ee
the TQ-equation and its dual both take the same ordinary form
\begin{align}
T Q'  &=  (u^{+})^{N} Q^{'--} + (u^{-})^{N} Q^{'++} \,, \non \\
T P'  &=  (u^{+})^{N} P^{'--} + (u^{-})^{N} P^{'++} \,,
\label{twistTQtransf}
\end{align}
except that $Q'$ and $P'$ are not polynomials.

Alternatively, one can transform only P
\be
P''(u) = e^{-u \phi} P(u) \,,
\label{twisttransf2}
\ee
in which case the Wronskian relation takes the ordinary form
\be
P^{''+} Q^{-} -  P^{''-} Q^{+}  = e^{-u \phi} u^{N} \,,
\label{twistWronsktransf}
\ee 
i.e. with trivial coefficients on the LHS, but $P''$ is not a polynomial.

\subsection{The open chain revisited}\label{sec:revisit}

Returning to the open spin chain, it is natural to ask whether
transformations analogous to (\ref{twisttransf1}) and
(\ref{twisttransf2}) can be found to bring the TQ-equations 
(\ref{TQrat}) and (\ref{TQratdual}) to the same more-ordinary form,
and to bring the Wronskian relation (\ref{Wronskrat}) to a more 
ordinary form, respectively.
As we shall see, only the latter is possible.

We begin by defining, in analogy with (\ref{twisttransf1}), the new Q-function 
\be
Q'(u) = S(u)\, Q(u) \,,
\label{S}
\ee
and we look for a function $S(u)$ that brings the TQ-equation to a 
more ordinary form. 
To this end, we multiply both sides of the TQ-equation (\ref{TQrat}) 
by $S$, and we demand that it have the same form as for the $SU(2)$-invariant 
open chain \cite{Bajnok:2019zub}
\be
- u\, T\, Q' = (u^{+})^{2N+1} Q^{'--} + (u^{-})^{2N+1} Q^{'++}\,,
\label{TQrattransf}
\ee
which requires that the function $S$ satisfy
\begin{align}
S^{--} &= g^{-}\, S\,, \label{S1} \\
S^{++} &= f^{+}\, S\,. \label{S2}
\end{align}
However, performing a $+$ shift of (\ref{S1}) and  $-$ shift of 
(\ref{S2}), we obtain the relations
\begin{align}
S^{-} &= g\, S^{+}\,, \non \\
S^{+} &= f\, S^{-}\,,
\end{align}
which imply the consistency condition $f g=1$, which is not satisfied. (Recall that $f$
and $g$ are given by (\ref{fgrat}).) We conclude that it is 
\emph{not} possible to bring the TQ-equation (\ref{TQrat}) to the 
more ordinary form (\ref{TQrattransf}) by the transformation (\ref{S}), and similarly for
the dual TQ-equation (\ref{TQratdual}).

We finally consider, in analogy with (\ref{twisttransf2}), the new P-function  
\be
P''(u) = V(u)\, P(u) \,,
\label{V}
\ee
and we look for a function $V(u)$ that brings the Wronskian relation 
to a more ordinary form. 
To this end, we multiply both sides of the Wronskian relation (\ref{Wronskrat}) 
by $V$, and we demand that it take the form
\be
P^{''+}\, Q^{-} - P^{''-}\, Q^{+} = u\, V\, Q_{0,0}\,,
\label{Wronskrattransf}
\ee
which requires that $P''$ satisfy
\begin{align}
P^{''+} &= g\, V\, P^{+} \,,  \label{V1} \\
P^{''-} &= f\, V\, P^{-} \,.    \label{V2}
\end{align}
Performing a $-$ shift of (\ref{V1}) and  $+$ shift of 
(\ref{V2}), we obtain the relations
\begin{align}
P^{''} &= g^{-}\, V^{-}\, P \,,  \non \\
P^{''} &= f^{+}\, V^{+}\, P \,,   
\end{align}
which imply that $V(u)$ must satisfy the functional relation
\be
\frac{V^{+}}{V^{-}} = \frac{g^{-}}{f^{+}}\,,
\ee
which has a solution in terms of products of gamma functions
\be
V(u) = \Gamma(i u + \alpha)\, \Gamma(-i u + \alpha)\, \Gamma(i u - 
\beta)\, \Gamma(-i u - \beta) \,.
\label{Gamma}
\ee
We conclude that the transformation (\ref{V}) with $V(u)$ given by 
(\ref{Gamma}) indeed brings the Wronskian relation (\ref{Wronskrat}) to the 
more ordinary form (\ref{Wronskrattransf}), where however $P''(u)$ is 
not polynomial.

% \bibliographystyle{utphys}
% \bibliography{refs}

\begin{thebibliography}{10}

\bibitem{Marboe:2016yyn}
C.~Marboe and D.~Volin, ``{Fast analytic solver of rational Bethe equations},''
  \href{http://dx.doi.org/10.1088/1751-8121/aa6b88}{{\em J. Phys.} {\bfseries
  A50} no.~20, (2017) 204002},
\href{http://arxiv.org/abs/1608.06504}{{\ttfamily arXiv:1608.06504 [math-ph]}}.
%%CITATION = ARXIV:1608.06504;%%.

\bibitem{Krichever:1996qd}
I.~Krichever, O.~Lipan, P.~Wiegmann, and A.~Zabrodin, ``{Quantum integrable
  systems and elliptic solutions of classical discrete nonlinear equations},''
  \href{http://dx.doi.org/10.1007/s002200050165}{{\em Commun. Math. Phys.}
  {\bfseries 188} (1997) 267--304},
\href{http://arxiv.org/abs/hep-th/9604080}{{\ttfamily arXiv:hep-th/9604080
  [hep-th]}}.
%%CITATION = HEP-TH/9604080;%%.

\bibitem{Kazakov:2007fy}
V.~Kazakov, A.~S. Sorin, and A.~Zabrodin, ``{Supersymmetric Bethe ansatz and
  Baxter equations from discrete Hirota dynamics},''
  \href{http://dx.doi.org/10.1016/j.nuclphysb.2007.06.025}{{\em Nucl. Phys.}
  {\bfseries B790} (2008) 345--413},
\href{http://arxiv.org/abs/hep-th/0703147}{{\ttfamily arXiv:hep-th/0703147
  [HEP-TH]}}.
%%CITATION = HEP-TH/0703147;%%.

\bibitem{Kuniba:2010ir}
A.~Kuniba, T.~Nakanishi, and J.~Suzuki, ``{T-systems and Y-systems in
  integrable systems},''
  \href{http://dx.doi.org/10.1088/1751-8113/44/10/103001}{{\em J. Phys.}
  {\bfseries A44} (2011) 103001},
\href{http://arxiv.org/abs/1010.1344}{{\ttfamily arXiv:1010.1344 [hep-th]}}.
%%CITATION = ARXIV:1010.1344;%%.

\bibitem{Marboe:2017dmb}
C.~Marboe and D.~Volin, ``{The full spectrum of AdS5/CFT4 I: Representation
  theory and one-loop Q-system},''
  \href{http://dx.doi.org/10.1088/1751-8121/aab34a}{{\em J. Phys.} {\bfseries
  A51} no.~16, (2018) 165401},
\href{http://arxiv.org/abs/1701.03704}{{\ttfamily arXiv:1701.03704 [hep-th]}}.
%%CITATION = ARXIV:1701.03704;%%.

\bibitem{Basso:2017khq}
B.~Basso, F.~Coronado, S.~Komatsu, H.~T. Lam, P.~Vieira, and D.-l. Zhong,
  ``{Asymptotic Four Point Functions},''
  \href{http://dx.doi.org/10.1007/JHEP07(2019)082}{{\em JHEP} {\bfseries 07}
  (2019) 082},
\href{http://arxiv.org/abs/1701.04462}{{\ttfamily arXiv:1701.04462 [hep-th]}}.
%%CITATION = ARXIV:1701.04462;%%.

\bibitem{Suzuki:2017ipd}
R.~Suzuki, ``{Refined Counting of Necklaces in One-loop $\mathcal{N}=4$ SYM},''
  \href{http://dx.doi.org/10.1007/JHEP06(2017)055}{{\em JHEP} {\bfseries 06}
  (2017) 055},
\href{http://arxiv.org/abs/1703.05798}{{\ttfamily arXiv:1703.05798 [hep-th]}}.
%%CITATION = ARXIV:1703.05798;%%.

\bibitem{Ryan:2018fyo}
P.~Ryan and D.~Volin, ``{Separated variables and wave functions for rational
  gl(N) spin chains in the companion twist frame},''
  \href{http://dx.doi.org/10.1063/1.5085387}{{\em J. Math. Phys.} {\bfseries
  60} no.~3, (2019) 032701},
\href{http://arxiv.org/abs/1810.10996}{{\ttfamily arXiv:1810.10996 [math-ph]}}.
%%CITATION = ARXIV:1810.10996;%%.

\bibitem{Coronado:2018ypq}
F.~Coronado, ``{Perturbative four-point functions in planar $ \mathcal{N}=4 $
  SYM from hexagonalization},''
  \href{http://dx.doi.org/10.1007/JHEP01(2019)056}{{\em JHEP} {\bfseries 01}
  (2019) 056},
\href{http://arxiv.org/abs/1811.00467}{{\ttfamily arXiv:1811.00467 [hep-th]}}.
%%CITATION = ARXIV:1811.00467;%%.

\bibitem{Jacobsen:2018pjt}
J.~L. Jacobsen, Y.~Jiang, and Y.~Zhang, ``{Torus partition function of the
  six-vertex model from algebraic geometry},''
  \href{http://dx.doi.org/10.1007/JHEP03(2019)152}{{\em JHEP} {\bfseries 03}
  (2019) 152},
\href{http://arxiv.org/abs/1812.00447}{{\ttfamily arXiv:1812.00447 [hep-th]}}.
%%CITATION = ARXIV:1812.00447;%%.

\bibitem{Bajnok:2019zub}
Z.~Bajnok, E.~Granet, J.~L. Jacobsen, and R.~I. Nepomechie, ``{On Generalized
  $Q$-systems},'' \href{http://dx.doi.org/10.1007/JHEP03(2020)177}{{\em JHEP}
  {\bfseries 03} (2020) 177},
\href{http://arxiv.org/abs/1910.07805}{{\ttfamily arXiv:1910.07805 [hep-th]}}.
%%CITATION = ARXIV:1910.07805;%%.

\bibitem{Sklyanin:1988yz}
E.~K. Sklyanin, ``{Boundary Conditions for Integrable Quantum Systems},''
\href{http://dx.doi.org/10.1088/0305-4470/21/10/015}{{\em J. Phys.} {\bfseries
  A21} (1988) 2375}.
%%CITATION = JPAGA,A21,2375;%%.

\bibitem{Gaudin:1971zza}
M.~Gaudin, ``{Boundary Energy of a Bose Gas in One Dimension},''
\href{http://dx.doi.org/10.1103/PhysRevA.4.386}{{\em Phys. Rev.} {\bfseries A4}
  (1971) 386--394}.
%%CITATION = PHRVA,A4,386;%%.

\bibitem{Alcaraz:1987uk}
F.~C. Alcaraz, M.~N. Barber, M.~T. Batchelor, R.~J. Baxter, and G.~R.~W.
  Quispel, ``{Surface Exponents of the Quantum XXZ, Ashkin-Teller and Potts
  Models},''
\href{http://dx.doi.org/10.1088/0305-4470/20/18/038}{{\em J. Phys.} {\bfseries
  A20} (1987) 6397}.
%%CITATION = JPAGA,A20,6397;%%.

\bibitem{Faddeev:1996iy}
L.~D. Faddeev, ``{How algebraic Bethe ansatz works for integrable models},'' in
  {\em Sym\'etries Quantiques (Les Houches Summer School Proceedings vol 64)},
  A.~Connes, K.~Gawedzki, and J.~Zinn-Justin, eds., pp.~149--219.
\newblock North Holland, 1998.
\newblock
\href{http://arxiv.org/abs/hep-th/9605187}{{\ttfamily arXiv:hep-th/9605187
  [hep-th]}}.
\newblock
%%CITATION = HEP-TH/9605187;%%.

\bibitem{Pronko:1998xa}
G.~P. Pronko and {\relax Yu}.~G. Stroganov, ``{Bethe equations 'on the wrong
  side of equator'},''
  \href{http://dx.doi.org/10.1088/0305-4470/32/12/007}{{\em J. Phys.}
  {\bfseries A32} (1999) 2333--2340},
\href{http://arxiv.org/abs/hep-th/9808153}{{\ttfamily arXiv:hep-th/9808153
  [hep-th]}}.
%%CITATION = HEP-TH/9808153;%%.

\bibitem{Mukhin:2009}
E.~Mukhin, V.~Tarasov, and A.~Varchenko, ``{Bethe algebra of homogeneous XXX
  Heisenberg model has simple spectrum},'' {\em Commun. Math. Phys.} {\bfseries
  288} (2009) 1--42, \href{http://arxiv.org/abs/0706.0688}{{\ttfamily
  arXiv:0706.0688 [math]}}.

\bibitem{Tarasov:2018}
V.~Tarasov, ``{Completeness of the Bethe ansatz for the periodic isotropic
  Heisenberg model},'' \href{http://dx.doi.org/10.1142/S0129055X18400184}{{\em
  Rev. Math. Phys.} {\bfseries 30} (2018) 1840018}.

\bibitem{Tsuboi:2009ud}
Z.~Tsuboi, ``{Solutions of the T-system and Baxter equations for supersymmetric
  spin chains},'' \href{http://dx.doi.org/10.1016/j.nuclphysb.2009.08.009}{{\em
  Nucl. Phys.} {\bfseries B826} (2010) 399--455},
\href{http://arxiv.org/abs/0906.2039}{{\ttfamily arXiv:0906.2039 [math-ph]}}.
%%CITATION = ARXIV:0906.2039;%%.

\bibitem{Tsuboi:2011iz}
Z.~Tsuboi, ``{Wronskian solutions of the T, Q and Y-systems related to infinite
  dimensional unitarizable modules of the general linear superalgebra
  $gl(M|N)$},'' \href{http://dx.doi.org/10.1016/j.nuclphysb.2013.01.007}{{\em
  Nucl. Phys.} {\bfseries B870} (2013) 92--137},
\href{http://arxiv.org/abs/1109.5524}{{\ttfamily arXiv:1109.5524 [hep-th]}}.
%%CITATION = ARXIV:1109.5524;%%.

\bibitem{Frassek:2015mra}
R.~Frassek and I.~M. Szecsenyi, ``{Q-operators for the open Heisenberg spin
  chain},'' \href{http://dx.doi.org/10.1016/j.nuclphysb.2015.10.010}{{\em Nucl.
  Phys.} {\bfseries B901} (2015) 229--248},
\href{http://arxiv.org/abs/1509.04867}{{\ttfamily arXiv:1509.04867 [math-ph]}}.
%%CITATION = ARXIV:1509.04867;%%.

\bibitem{Baseilhac:2017hoz}
P.~Baseilhac and Z.~Tsuboi, ``{Asymptotic representations of augmented
  q-Onsager algebra and boundary K-operators related to Baxter Q-operators},''
  \href{http://dx.doi.org/10.1016/j.nuclphysb.2018.02.017}{{\em Nucl. Phys.}
  {\bfseries B929} (2018) 397--437},
\href{http://arxiv.org/abs/1707.04574}{{\ttfamily arXiv:1707.04574 [math-ph]}}.
%%CITATION = ARXIV:1707.04574;%%.

\bibitem{Ghoshal:1993tm}
S.~Ghoshal and A.~B. Zamolodchikov, ``{Boundary S matrix and boundary state in
  two-dimensional integrable quantum field theory},''
  \href{http://dx.doi.org/10.1142/S0217751X94001552}{{\em Int. J. Mod. Phys.}
  {\bfseries A9} (1994) 3841--3886},
  \href{http://arxiv.org/abs/hep-th/9306002}{{\ttfamily arXiv:hep-th/9306002
  [hep-th]}}.
[Erratum: Int. J. Mod. Phys.A9,4353 (1994)].
%%CITATION = HEP-TH/9306002;%%.

\bibitem{deVega:1993xi}
H.~J. de~Vega and A.~Gonzalez-Ruiz, ``{Boundary K matrices for the XYZ, XXZ and
  XXX spin chains},'' \href{http://dx.doi.org/10.1088/0305-4470/27/18/021}{{\em
  J. Phys.} {\bfseries A27} (1994) 6129--6138},
\href{http://arxiv.org/abs/hep-th/9306089}{{\ttfamily arXiv:hep-th/9306089
  [hep-th]}}.
%%CITATION = HEP-TH/9306089;%%.

\bibitem{Cao:2013qxa}
J.~Cao, W.-L. Yang, K.~Shi, and Y.~Wang, ``{Off-diagonal Bethe ansatz solution
  of the XXX spin-chain with arbitrary boundary conditions},''
  \href{http://dx.doi.org/10.1016/j.nuclphysb.2013.06.022}{{\em Nucl. Phys.}
  {\bfseries B875} (2013) 152--165},
\href{http://arxiv.org/abs/1306.1742}{{\ttfamily arXiv:1306.1742 [math-ph]}}.
%%CITATION = ARXIV:1306.1742;%%.

\bibitem{Nepomechie:2013ila}
R.~I. Nepomechie, ``{An inhomogeneous T-Q equation for the open XXX chain with
  general boundary terms: completeness and arbitrary spin},''
  \href{http://dx.doi.org/10.1088/1751-8113/46/44/442002}{{\em J. Phys.}
  {\bfseries A46} (2013) 442002},
\href{http://arxiv.org/abs/1307.5049}{{\ttfamily arXiv:1307.5049 [math-ph]}}.
%%CITATION = ARXIV:1307.5049;%%.

\bibitem{Wang2015}
Y.~Wang, W.-L. Yang, J.~Cao, and K.~Shi, {\em Off-Diagonal Bethe Ansatz for
  Exactly Solvable Models}.
\newblock Springer, 2015.

\bibitem{Nepomechie:2017hgw}
R.~I. Nepomechie, R.~A. Pimenta, and A.~L. Retore, ``{The integrable quantum
  group invariant $A_{2n-1}^{(2)}$ and $D_{n+1}^{(2)}$ open spin chains},''
  {\em Nucl. Phys.} {\bfseries B924} (2017) 86--127,
\href{http://arxiv.org/abs/1707.09260}{{\ttfamily arXiv:1707.09260 [math-ph]}}.
%%CITATION = ARXIV:1707.09260;%%.

\end{thebibliography}

\providecommand{\href}[2]{#2}\begingroup\raggedright\endgroup

\end{document}